\documentclass[preprint,showpacs,showkeys,preprintnumbers,aps]{revtex4}

\usepackage{graphicx}

\begin{document}

\preprint{Fecher et al., Bulk sensitive photo emission spectroscopy of $C1_b$~compounds.}

\title{Bulk sensitive photo emission spectroscopy of $C1_b$~compounds.}

\author{Gerhard H. Fecher}
\email{fecher@uni-mainz.de}
\author{Andrei Gloskowskii}
\author{Kristian Kroth}
\author{Joachim Barth}
\author{Benjamin Balke}
\author{Claudia Felser}
\affiliation{Institut f\"ur Anorganische Chemie und Analytische Chemie,\\
         Johannes Gutenberg~-~Universit\"at,
         55099~Mainz, Germany}

\author{Franz Sch{\"a}fers}
\author{Marcel Mertin},
\author{Wolfgang Eberhardt}
\affiliation{BESSY GmbH, Albert-Einstein-Stra{\ss}e~15, 12489 Berlin, Germany}

\author{Sven M{\"a}hl,}
\author{Oliver Schaff}
\affiliation{SPECS GmbH, Voltastra{\ss}e~5, 13355~Berlin, Germany}
\date{\today}

\begin{abstract}

This work reports about bulk-sensitive, high energy photoelectron
spectroscopy from the valence band of CoTiSb excited by photons from
1.2 to 5~keV energy. The high energy photoelectron spectra were taken
at the KMC-1 high energy beamline of BESSY II employing the recently
developed {\scshape Phoibos} 225 HV analyser. The measurements show a
good agreement to calculations of the electronic structure using the
LDA scheme. It is shown that the high energy spectra reveal the bulk
electronic structure better compared to low energy XPS spectra.

\end{abstract}



\maketitle

\section{Introduction}

Photo emission spectroscopy is the method of choice to study the
occupied electronic structure of materials. Low kinetic energies result
in a low electron mean free path being only 5.3~{\AA} at kinetic
energies of 100~eV or 26{\AA} at 1.2~keV (all values calculated for
CoTiSb using the Tamuna-Powel-Penn (TPP-2M) equations \cite{TPP93}). A
depth of less than one cubic $C1_b$ cell will contribute to the
observed intensity if using VUV ($<40$~eV) or XUV ($<120$~eV) light for
excitation. The situation becomes much better at high energies. In the
hard X-ray region of about 5~keV one will reach a high bulk sensitivity
with an escape depth being larger than 84~{\AA} (corresponding to 15
cubic cells). Lindau {\it et al} \cite{LPD74} demonstrated in 1974 the
possibility of high energy photo emission with energies up to 8~keV,
however, no further attention was devoted to such experiments for many
years. High energy photo emission (at about 15~keV excitation energy)
was also performed as early as 1989 \cite{Mei89} using a $^{57}$Co
M{\"o\ss}bauer $\gamma$-source for excitation, however, with very low
resolution only. Nowadays, high energy excitation and analysis of the
electrons become easily feasible due to the development of high intense
sources (insertion devices at synchrotron facilities) and multi-channel
electron detection. Thus, high energy X-ray photo emission spectroscopy
(HXPS) was recently introduced by several groups
\cite{KYT03,SS04,TKC04,Kob05,PCC05,TSC05} as a bulk sensitive probe of
the electronic structure in complex materials. In the present work,
excitation energies of $h\nu\ = 1.2\ldots 5$~keV were used to study the
density of states of CoTiSb.

CoTiSb crystallises in the $C1_b$ structure. This structure type is
often observed in ternary (XYZ) transition metal (X, Y) intermetallics
with a main group element (Z). It is based on a face-centered cubic
lattice (space group $F\:\overline{4}3m$) similar to the binary
zinc-blende type semiconductors. Many of the $C1_b$ compounds belong to
the class of half-metallic ferromagnets \cite{Gro83}. CoTiSb carries no
magnetic moment according to the Slater-Pauling rule because it has
overall 18 valence electrons. Most of the $C1_b$ compounds with 18
valence electrons are found to be semi-conducting \cite{Tob98}.

\section{Experimental and calculational details}
\label{sec:ED}

CoTiSb samples have been prepared by arc melting of stoichiometric
amounts of the constituents in an argon atmosphere at 10$^{-4}$~mbar.
Care has been taken to avoid oxygen contamination. This was ensured by
evaporating Ti inside of the vacuum chamber before melting the compound
as well as by additional purification of the process gas. After cooling
of the resulting polycrystalline ingots, they were annealed in an
evacuated quartz tube for 21~days. This procedure resulted in samples
exhibiting the $C1_b$ structure, which was verified by X-ray powder
diffraction (XRD). Magneto-structural investigations on Fe substituted
samples (CoTi$_{1-x}$Fe$_x$Sb) were carried out using $^{57}$Fe
M{\"o\ss}bauer spectroscopy. The magnetic properties of Fe doped
samples were investigated by a super conducting quantum interference
device. The resistivity of the samples was measured by a 4-point probe.
Further details and results of the structural and magnetic properties
are reported elsewhere \cite{BKF06}.

An ESCALAB MkII~-~HV (VG) for kinetic energies up to 15~keV
\cite{Mei89,KMG93} has been used here to take valence band spectra at
an excitation energy of 1.253~keV (Mg~K$_\alpha$ with a line width of
$\Delta E = xx$~eV) for comparison. For this purpose, the slits and
pass energy have been set for a resolution of 200~meV. The energy is
calibrated at the Au 4f$_{7/2}$ emission line. X-ray photoelectron
spectroscopy (XPS) was also used to verify the composition and to check
the cleanliness of the samples. After removal of the native oxide from
the polished surfaces by Ar$^+$ ion bombardment, no impurities were
detected with XPS.

\begin{figure}
\includegraphics[width=11cm]{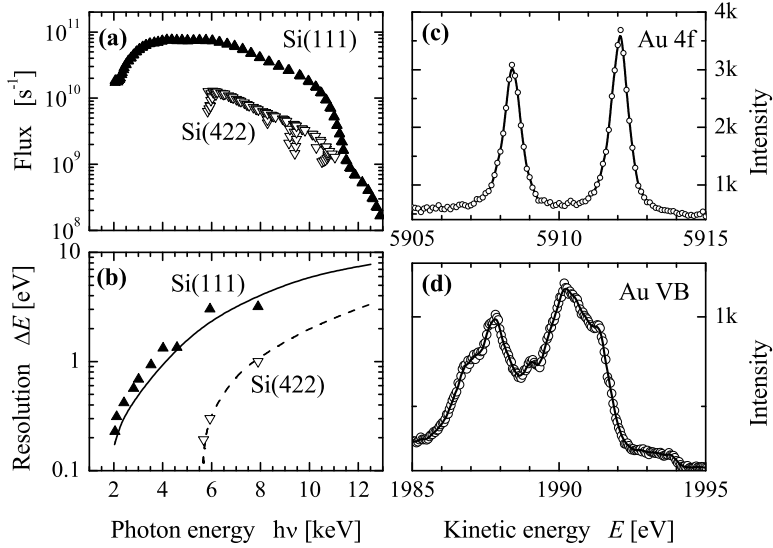}
\caption{Characteristics of the KMC-1 monochromator and the {\scshape Phoibos} 225 HV electron analyser. \newline
         (a) shows the flux normalised to a current of 100~mA in the storage ring and
         (b)~the calculated (lines) and measured (symbols) resolution of the monochromator.
         (c)~and~(d) show typical spectra of the Au 4f core level and valence band
         excited by 6~keV (Si(422))and 2~keV (Si(111)) photons, respectively.}
\label{fig_1}
\end{figure}

Details of the electronic structure have been explored experimentally
by means of high energy X-ray photo emission spectroscopy. The
measurements have been performed at the KMC-1 beamline of the storage
ring BESSY II (Berlin, Germany). The photons are produced by means of a
bending magnet. The photon beam is focused by a toroidal mirror and
monochromatised by a double-crystal monochromator. Three different
crystal pairs -~that are InSb(111), Si(111), and Si(422)~- can be
employed for photon energies up to $\approx 15$~keV with starting
energies of 1.674, 1.997, and 5.639~keV, respectively. The crystals can
be changed in situ. The acceptance of the toroidal Si/Pt mirror is $(6
\times 0.5)$~mrad$^2$. The resulting spot size of the beam is $(0.4
\times 0.6)$~mm$^2$ at the focus being located about 35~m behind the
source. The complete beamline is operated under oil-free UHV conditions
($<5 \times 10^{-8}$~mbar at last valve). The characteristics of the
beamline -~flux and resolution using Si(111) and Si(422) crystals~- are
displayed in Figure~\ref{fig_1}~(a) and~(b).

As electron analyser, a recently developed hemispherical spectrometer
with 225~mm radius has been used ({\scshape Phoibos} 225 HV). The
analyser is designed for high resolution spectroscopy at kinetic
energies up to 15~keV. For the present study, a 2D-CCD detector has
been employed for detection of the electrons. The analyser is prepared
for parallel 2D (energy and angle) and simultaneous spin-detection
using a combination of a 2D-delay-line detector with a low energy
Mott-detector. Typical spectra, as taken from a Au(100) single crystal
are shown in Figure~\ref{fig_1}~(c) and~(d).

Under the present experimental conditions an overall resolution of
240~meV at 2~keV photon energy has been reached (monochromator plus
electron detector), as was determined from the Fermi edge of Au(100).
Due to the low cross-section of the valence states from the
investigated compounds, the spectra had to be taken with $E_{pass}$
from 50~eV to 150~eV and a 1~mm entrance slit for a good signal to
noise ratio. The polycrystalline CoTiSb samples have been cleaned
in-situ by Ar$^+$ ion bombardment before taking the spectra to remove
the native oxide layer. Core-level spectra have been taken to check the
cleanliness of the samples. No traces of impurities were found. All
measurements have been taken at room temperature ($\approx 300$~K).

The self-consistent electronic structure calculations have been carried
out using the scalar-relativistic full potential linearised augmented
plane wave method (FLAPW) as provided by Wien2k \cite{BSM01}. In the
parametrisation of Perdew {\it et al} \cite{PCV92} the
exchange-correlation functional was taken within the generalised
gradient approximation (GGA). A $25\times25\times25$ mesh has been used
for integration, resulting in 819 $k$-points in the irreducible wedge
of the Brillouin zone. The properties of CoTiSb have been calculated in
$F\:\overline{4}3m$ symmetry using the experimental lattice parameter
($a=5.883$~{\AA}) as determined by XRD. Co atoms are placed on 4a
Wyckoff positions, Ti on 4c and Sb on 4d. This leaves the vacancy on
the 4b position in accordance with the Rietveldt refinement of the XRD
data. All muffin tin radii have been set as nearly touching spheres
with $r_{MT}=2.39 a_{0B}$ for the 3d elements and $2.25 a_{0B}$ for the
main group element ($a_{0B} = 0.529177$~{\AA}). A structural
optimisation for the compound showed that the calculated lattice
parameter deviates from the experimental one only marginally.

\section{Results and Discussion}

\begin{figure}
\includegraphics[width=7cm]{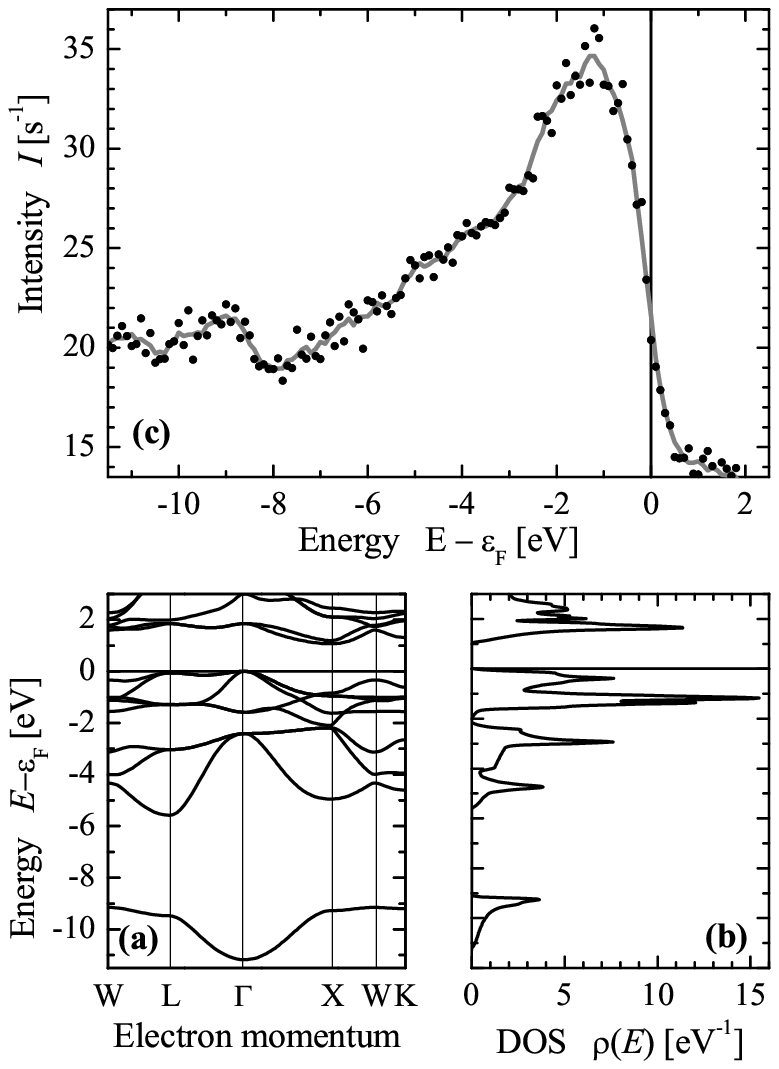}
\caption{XPS and density of states of CoTiSb. \newline
         Shown are the band structure (a) and the density of states (b)
         in comparison to a XPS spectrum excited by Mg~K$_\alpha$ radiation (c).}
\label{fig_2}
\end{figure}

\begin{figure}
\includegraphics[width=11cm]{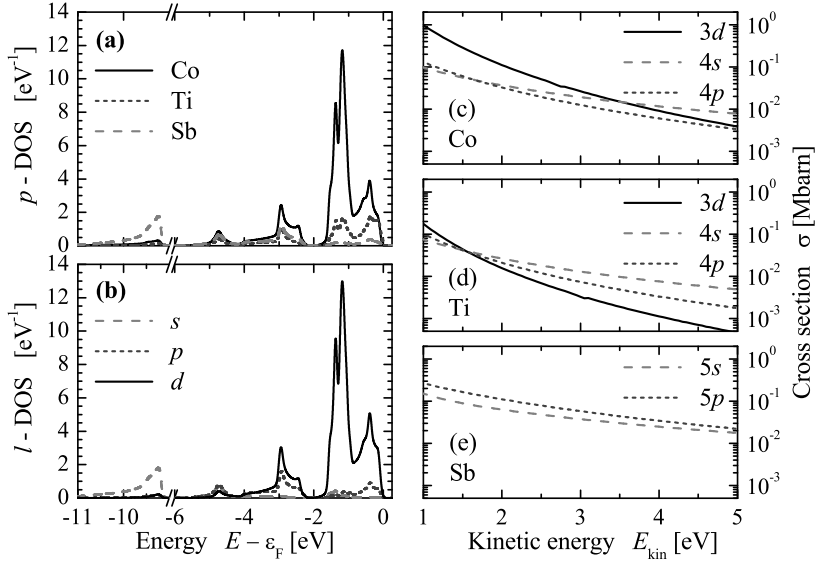}
\caption{Partial and orbital resolved densities of states and photo emission cross sections. \newline
         Shown are the atomic resolved $p$-DOS in (a)
         and the orbital momentum resolved $l$-DOS in (b).
         (Note the axis break between -9~eV and -6~eV that was used to bridge
         the low lying hybridisation gap.) \newline
         Panels (c)-(e) show the calculated energy and $l$ dependence of the cross section
         of the valence states of atomic Co, Ti, and Sb.}
\label{fig_3}
\end{figure}

Figures \ref{fig_2}~(a) and~(b) display the calculated band structure
and density of states (DOS). The compound exhibits a clear gap in the
band structure and DOS, that means it is semiconductor. The size of the
gap amounts to about 1.45~eV. It is clearly visible that the gap is
surrounded by regions of high density emerging from flat $d$-bands. The
high density in the valence band emerges mainly from states located at
the Co site, whereas the high density at the bottom of the conduction
band is related to states located at the Ti site. The gap is a result
of hybridisation with Sb states. The states at about 10eV below the
Fermi energy ($\epsilon_F$) are mainly located at the Sb site and have
$s$ like character (see Fig.\ref{fig_3}). This behaviour is typical for
$C1_b$ compounds with overall 18 valence electrons in the unit cell.

The partial, atomic resolved ($p$) and the orbital ($l$) resolved
densities of states are shown in Figure \ref{fig_3} (a) and (b),
respectively. The $p$-DOS of the interstitial as well as the $l$-DOS
for higher angular momenta ($l$) are omitted as they contribute only
very few to the total density of states. The main contribution to
states close to the Fermi energy are $d$-states being located at the Co
site. As typical for $C1_b$ compounds with $T_d$ symmetry, one finds a
strong bonding interaction between transition metal $d$-states
($t_{2g}$) with Sb $p$-states ($t_{1u}$) that hybridise mainly along
the $\Gamma-L$ direction \cite{JKW00}. This hybridisation leads here to
the peak in the density at about 3~eV below $\epsilon_F$. The smaller
peak at about -5~eV is due to $s-p$ hybridisation.

The valence band spectrum excited by Mg~K$_{\alpha}$ radiation is shown
in Figure~\ref{fig_2}~(c). On a first sight, the spectrum exhibits a
high intensity close to the Fermi energy and is rather unstructured at
higher binding energies. The low lying $s$ band at 9~eV to 11~eV below
$\epsilon_F$ is only weakly revealed. Besides the maximum, there are
only weak structures (compared to the calculated DOS) detectable in the
energy region from -6~eV to $\epsilon_F$. The maximum is located about
1.4~eV below $\epsilon_F$ and corresponds to the high density of Co
$d$-states seen in the DOS.

\begin{figure}
\includegraphics[width=7cm]{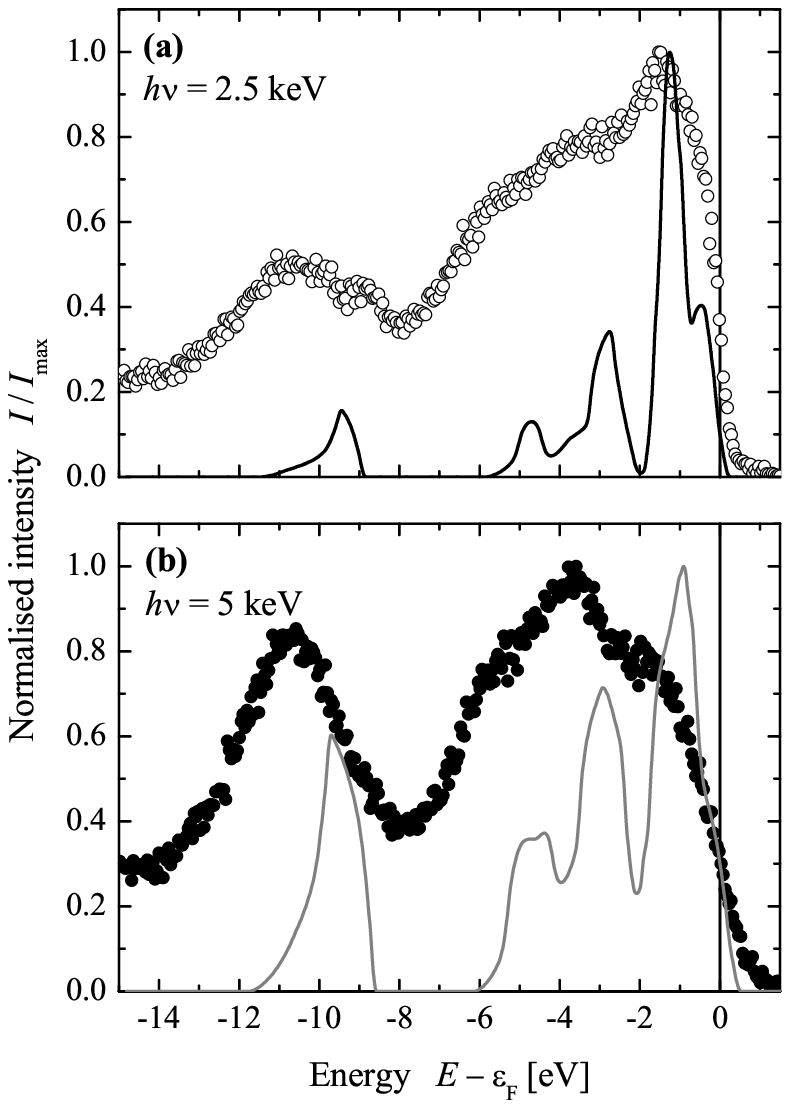}
\caption{High energy valence band spectra of CoTiSb. \newline
         Shown are spectra taken at 2.5~keV (a) and 5~keV (b) excitation energy (symbols)
         and the DOS weighted by the photo emission cross section (lines).
         }
\label{fig_4}
\end{figure}

The results from high energy photo emission are shown in
Figure~\ref{fig_4} and compared to the total density of states weighted
by the partial cross sections. For better comparison, the weighted DOS
is additionally broadened by a Gaussian of 0.25~eV (0.5~eV) width to
account roughly for the experimental resolution at 2.5~keV (5~keV)
excitation energy.

Both high energy spectra reveal clearly the low lying $s$-states at
about -11~eV to -9~eV below the Fermi energy, in well agreement to the
calculated DOS. The intensity in this energy region is considerably
higher compared to the excitation at 1.2~keV (see Fig.~\ref{fig_2}).
These low lying bands are separated from the high lying $d$-states by
the $C1_b$-typical hybridisation gap being clearly resolved in the
spectra as well as the calculated DOS. The size of this gap amounts
typically to $\Delta E\approx 3\ldots 4$~eV in Sb containing compounds.

Obviously, the emission from the low lying $s$-states is pronouncedly
enhanced compared to the emission from the $d$-states. This can be
explained by a different behaviour of the cross sections of the $s$,
$p$, and $d$ states with increasing kinetic energy as was recently
demonstrated by Panaccione {\it et al} for the case of the silver
valence band \cite{PCC05}. The orbital momentum and site resolved cross
sections are displayed in Fig.~\ref{fig_3}~(c)-(e). The calculations
were performed for atomic valence states using a modified full
relativistic Dirac-solver based on the computer programs of Salvat and
Mayol \cite{SMa91,SMa93}. The radial integrals for the various
transitions have been computed using the dipole length-form. In
particular, the cross section for $d$-states decreases faster with
increasing photon energy than the ones of the $s$ or $p$-states. This
behaviour influences also the onset of the $d$-bands at about -7~eV to
-6~eV. Just at the bottom of those $d$-bands, they are hybridised with
$sp$-like states, leading to a high intensity in this energy region. At
$h\nu=2.5$~keV, the structures of the weighted DOS have approximately
the same heights compared to the DOS shown in Fig.~\ref{fig_2}~(b),
whereas the strong enhancement of the $s$ states is clearly visible if
comparing Fig.~\ref{fig_4}~(a) and~(b).

The structure of the high energy spectra in the range of the $d$ states
agrees roughly with the structures observed in the total DOS although
the high density at -5~eV and -3~eV is not well resolved. Overall the
emission from the $d$ states covers a larger energy range compared to
the calculated DOS, what gives advice on an underestimation of
correlation effects in the local density approximation. However, one
also has to account for lifetime broadening and the experimental
resolution if comparing that energy range. At 2.5~keV excitation, the
emission is still dominated by the high dense $d$-states at about
-1.5~eV. Increasing the excitation energy to 5~keV (note the lower
resolution of the monochromator at that energy) has the result that the
intensity in this energy range becomes considerably lower. At the same
time, the emission from the bands at about -3~eV becomes strongly
enhanced. As both structures in the DOS emerge, at least partially,
from flat $d$-bands, the transfer of the intensity maximum might be not
only explained by pure cross section effects.

The valence band spectra may be seen as a convolution of the initial
and final state DOS. The final state DOS is rather constant at high
kinetic energies and final state effects may play a minor rule only.
Two weighting factors enter the DOS-convolution. The first is the
transition matrix element that contains both the selection rules and
the cross sections (radial matrix elements). The radial matrix elements
are partially responsible for the rearrangement of the orbital resolved
intensities as discussed above. The second weighting factor is the
complex self-energy of the photoelectron. Among other things, it
depends on the hole lifetime. At low kinetic energies, the spectra are
obviously governed by the long life time of the holes at binding
energies close to $\epsilon_F$. At high kinetic energies, where the
sudden approximation is reached, the photoelectron is not as strongly
coupled to its hole and the lifetime at $\epsilon_F$ plays less a role.
Not only from the mean free path but also from the presented point of
view, the high energy photo emission will help to understand the bulk
electronic structure better than using only low energy XPS.

Overall, the measured photoelectron spectra agree with the calculated
density of states. Small shifts of the peaks in the measured spectra
compared to the calculated DOS give advice on an incomplete treatment
of correlation effects in the local density approximation.

\section{Summary and Conclusions}

The electronic structure of the ternary $C1_b$ compound CoTiSb was
investigated by means of XPS. True bulk sensitive, high energy photo
emission indicates that the cross section differs not only between
states of different angular momentum but also depends on the binding
energy for a fixed angular momentum. The first effect leads to a
pronounced emission from low lying $s$ bands. The second effects leads
to a pronounced transfer of intensity away from the Fermi energy. It
can be explained by a reduction of life time effects. Those lifetime
effects govern the spectrum at low excitation energy and lead to an
enhanced intensity close to the Fermi energy at low kinetic energies.
Such lifetime effects can be reduced by use of high energy photo
emission.

\begin{acknowledgments}
Financial support by the Deutsche Forschungs
Gemeinschaft (project TP7 in research group FG 559) as well as by BESSY is gratefully acknowledged.
\end{acknowledgments}

\bibliography{Fecher_ICESS10_cm}

\end{document}